# A weak equivalence principle test on a suborbital rocket

Robert D. Reasenberg and James D. Phillips
Smithsonian Astrophysical Observatory
Harvard-Smithsonian Center for Astrophysics

E-mail: reasenberg@cfa.harvard.edu

Abstract. We describe a Galilean test of the weak equivalence principle, to be conducted during the free fall portion of a sounding rocket flight. The test of a single pair of substances is aimed at a measurement uncertainty of $\sigma(\eta) < 10^{-16}$ after averaging the results of eight separate drops. The weak equivalence principle measurement is made with a set of four laser gauges that are expected to achieve $0.1$ pm $Hz^{-1/2}$. The discovery of a violation ($\eta \neq 0$) would have profound implications for physics, astrophysics, and cosmology.

## 1. Introduction

The weak equivalence principle (WEP) underlies general relativity, but violations are possible in most theories being developed to unify gravity with the other forces. A WEP violation is characterized by $\eta$, which is identically zero in any metric theory of gravity, including general relativity [e.g., Fischbach and Talmadge 1999],

$$\eta_{AB} = \frac{a_A - a_B}{(a_A + a_B)/2}$$

where $a_A$ and $a_B$ are the acceleration of bodies A and B, these bodies are moving under the influence of identical gravity fields, and there is no other cause of the acceleration. The discovery of a WEP violation would have profound implications for physics, astrophysics and cosmology.

The present best tests of the WEP are made using a rotating torsion pendulum and yield $\sigma(\eta) < 2 \; 10^{-13}$ [Schlamminger et al. 2008]. Advancement of this approach has slowed because of intrinsic problems with the suspension fiber that may be overcome by operating at LHe temperature [Newman 2001, Berg et al. 2005]. LHe temperature torsion balances are also being used to study short range gravity [Hammond 2007]. There are several proposals for better WEP tests in an Earth orbiting spacecraft [Sumner 2007, Nobili et al. 2009, Chhun et al 2007]. The most sensitive among these is the satellite test of the equivalence principle (STEP) [Overduin et al 2009]. This cryogenic experiment is based on technological heritage from the successful GPB mission [GP-B]. STEP is aimed at a measurement uncertainty of $\sigma(\eta) < 10^{-18}$.

## 2. Instrument Design and Operation

We expect that SR-POEM, the sounding rocket version of POEM (Principle Of Equivalence Measurement) will achieve a sensitivity $\sigma(\eta) < 10^{-16}$ in the ~10 minute science portion of a single flight of a suborbital rocket. Here, we describe SR-POEM in terms of the present straw-man design, which is an outgrowth of the ground-based version (POEM) [Reasenberg and Phillips 2007]. In the sounding rocket payload, we compare the rate of fall of two test substances, "A" and "B."  Correspondingly, the instrument carries two test mass assemblies, TMA-A and TMA-B.  There has been much discussion of the "best choices" of substances [e.g., Blaser 1996] and the strategy for selecting several pairs for an experiment.  The nominal pair for the first flight is Al & Pb.

Each TMA comprises a pair of aluminum cubes connected by a short tube or pair of tubes (Fig. 1).  In one TMA, cylinders of Al have been removed and replaced by tubes of Pb without changing the TMA mass and having minimal effect on the moments of inertia.  The cubes are arranged in a square lying in a plane perpendicular to the fore-aft axis of the payload (the z axis) and close to the CM of the free-flying payload.  The two cubes along a diagonal of the square have the same kind of sample and are joined.  At a distance of 0.3 m along the z axis, there is a highly stable ULE glass plate that holds four of our tracking frequency laser gauges (TFG) [Phillips and Reasenberg 2005] with the same in-plane spacing and orientation as the cubes.  As in POEM, the cubes are surrounded by the plates of a multi-component capacitance gauge.  The SR-POEM capacitance gauge senses all six degrees of freedom for both cubes of each TMA, providing redundant information.  The capacitance gauge plates can be used to apply force or torque electrostatically.  This capability, when combined with the sensing, is called the TMA suspension system (TMA-SS).  In operation, each cube is continuously observed by the corresponding TFG.

NASA's Black Brant XI, our nominal launch vehicle, achieves an altitude of ~1200 km with a 160 kg payload.  When it reaches an altitude of ~800 km, the science phase of the flight begins.  By that time, the starting conditions for the first drop have been achieved:  The TMA have been uncaged, the z axis has been aligned with the direction to the Earth's center of mass[1], the TMA charges have been neutralized, and the TMA-SS has been used to assess and then correct the position and motion of the TMA in all six degrees of freedom.

At the start of a drop, the TMA-SS is disabled after which data taken by the TFG may be used for a WEP measurement lasting $Q$ (= 40 s, see below).  There will be eight such measurements during the science phase of the flight, with the instrument orientation reversed between successive measurements to cancel most systematic errors.  This "payload inversion" has the effect of reversing the WEP violating signal in the instrument coordinate system.  The gravity of the payload is unaffected by the inversion; it produces the same acceleration of the TMA in the instrument coordinate system independent of payload orientation.  During the payload-inverting rotational slews, the TMA-SS again controls each TMA's six degrees of freedom.

Finally, TFG data from the measurement cycles will be combined in a weighted-least-squares fit to estimate η and its formal uncertainty.  Required auxiliary data will include the payload trajectory from an on-board GPS receiver,  payload attitude from the spacecraft attitude control system (ACS), TMA transverse velocities from the capacitance gauge, and an Earth gravity model.

## 2.1. EP measurement sensitivity.

If a single laser gauge has distance measurement uncertainty $\sigma_0$ for a sampling time $\tau_0$, then the uncertainty in the estimate of acceleration, based on many measurements (assuming white noise) with total time $T$, is

$$\sigma_{acc}(T) = \frac{\sigma_0(\tau_0)}{Q^2} \sqrt{\frac{\tau_0}{T}} K ,$$

where $Q$ is the free-fall time, and the value of $K$ is obtained from an analytic covariance analysis in which the initial position and velocity must be estimated along with the acceleration to obtain an unbiased acceleration estimate: $K = 12\sqrt{5} \approx 27$. The corresponding WEP sensitivity for the pair of TMA and four TFG is $\sigma(\eta) = \sigma_{acc}/(R\,g(h))$, where $R$ is the fraction of TMA mass that is sample (test substance), and $g(h)$ is the acceleration of gravity at the altitude $h$ of the instrument.

---

[1] We are still investigating the orientation of the payload during a measurement: inertial, with z axis along the mean nadir direction; or rotating at a fixed rate, with the z axis pointing toward nadir as closely as possible.  (Because of the trajectory, a uniform rotation rate cannot be continuously nadir pointing.)

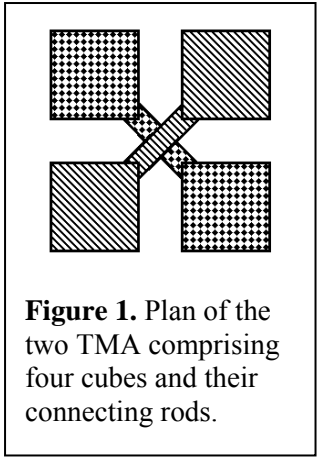

**Figure 1.** Plan of the two TMA comprising four cubes and their connecting rods.

In a reasonable scenario, $\sigma_o(1s) = 0.1$ pm, $Q = 40$ s, $R = 0.5$, and there are eight measurement cycles so that $T = 320$ s. Then, $\sigma(\eta) = 0.2 \times 10^{-16}$ before allowance for uncorrected sources of systematic error. We plan to have the eight measurements placed symmetrically around the payload apogee. Following a payload reorientation, each TMA is brought back to its nominal starting position by the TMA-SS before that system is again disabled. Assuming it takes 10 s for the suspension system to act and the TMA to stabilize after the payload reorientation, the measurement sequence, including the seven reorientations of about 20 s each, would take $8 \times 40 + 7 \times (20 + 10) = 530$ s and occupy the top 300 km of the trajectory. We would like to be above 800 km to suppress air drag acceleration of the spacecraft. Thus, in the present scenario, the SR-POEM experiment would fit within the capability of the NASA Sounding Rocket Program.

*2.2. Instrument Design.*

The principal driver of the instrument design is the prevention of systematic error. The CM positions of the two TMA are nominally the same, as described below. With the arrangement shown in Fig. 1, the z coordinate of a TMA CM is obtained from the average of the two TFG measurements made to the mirrors on the two cubes.

The instrument precision structure is constructed of ULE glass, which has a low coefficient of thermal expansion (CTE, about $10^{-8}$ / K). This structure has three major components: 1) The TMA plate holds capacitance gauge electrode sets and provides access for the TMA caging mechanism. The electrode sets are envisioned as being ULE or quartz boxes with electrodes deposited on the inside.[2] They require precision (10 micron) placement on the TMA plate. 2) The metering structure provides a rigid, stable connection between the TMA plate and the TFG plate. It will be joined to the TMA plate most likely by either frit bonding or KOH bonding. 3) The TFG plate supports the four TFG, the associated beam steering devices, and the optical detectors. All components are on the fore side (top in Fig. 2) of the plate, and the beam launch optics are mounted over holes in the plate to allow a view of the TMA. The TFG plate is connected to the metering structure by means of three flexures, each constraining two degrees of freedom, and likely made of titanium. (See, for example, the Itek flexure design in Fig. 3 of Reasenberg et al. [1996].) This design is intended to impose a minimum time-varying stress on the TFG plate and thus minimize changing plate distortion.

The measurement system is inside a vacuum chamber that is pumped down, baked and allowed to cool before launch. A small pump runs during the flight. The prime candidates are an ion pump and a sorption pump. A flange at the aft end of the chamber supports the principal components by means of four flexures (two flat blade and two stingers) that define the six degrees of freedom of the support ring, yet maintain a four-fold symmetry in the thermal connection. That ring is connected to the precision structure by means of four low conductivity (Ti) screws and thermally isolating spacers, Fig. 2. All electrical and optical connections to the instrument are made through this flange. At the other end, the chamber has a welded dome with two flanges for pumping: one for the pump that will run during the

---

[2] For example, the plates that will be assembled into the box could be metalized by vapor deposition, and the gaps between the electrodes etched using photoresist-based techniques.

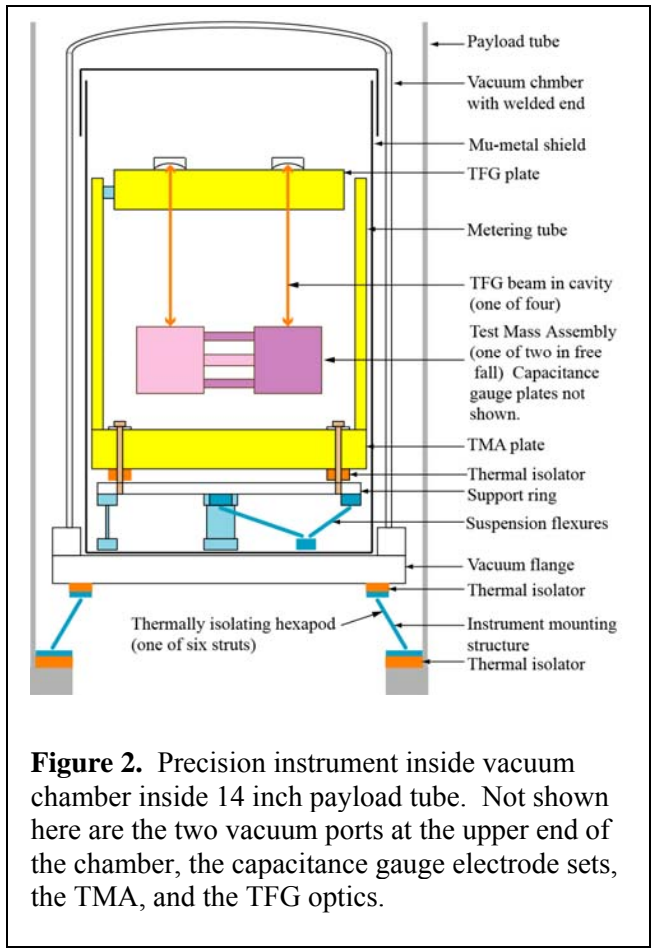

**Figure 2.** Precision instrument inside vacuum chamber inside 14 inch payload tube. Not shown here are the two vacuum ports at the upper end of the chamber, the capacitance gauge electrode sets, the TMA, and the TFG optics.

*Labels (top to bottom):* Payload tube; Vacuum chmber with welded end; Mu-metal shield; TFG plate; Metering tube; TFG beam in cavity (one of four); Test Mass Assembly (one of two in free fall) Capacitance gauge plates not shown.; TMA plate; Thermal isolator; Support ring; Suspension flexures; Vacuum flange; Thermal isolator; Thermally isolating hexapod (one of six struts); Instrument mounting structure; Thermal isolator flight, and one for a valve that will allow the chamber to be connected to a pumping station to initially prepare the vacuum, nominally to $10^{-8}$ Torr.

The experiment, i.e., the vacuum chamber and its contents, is connected to the payload tube through a thermally isolating hexapod motion system (Stewart platform). By means of control of the hexapod, the experiment becomes effectively drag free and its rotational fluctuations are significantly reduced.

### 2.3. Laser gauge.

The principal measurements are made by the four TFG observing the four cubes that are joined to form the two TMA. Because of its high precision, 0.1 pm in 1 s, the TFG allows a high accuracy test of the WEP during the brief period afforded by a sounding rocket. The TFG speed of measurement, coupled with the payload inversions, shifts upward the frequency band of interest for the measurements to 0.007 Hz and above. This shift reduces susceptibility to systematic error.

The TFG [Noecker et al. 1993, Reasenberg et al. 1995] was developed by us for POINTS, a low-cost spaceborne astrometric optical interferometer [Reasenberg et al. 1992 and references therein]. The TFG has five principal advantages for SR-POEM. 1. It is intrinsically free of the nm-scale cyclic bias characteristic of the conventional, heterodyne laser gauge (e.g., Hewlett-Packard, Zygo). 2. It uses one beam, not two, which simplifies the beam launcher. 3. The distance changes are translated to a radio frequency signal, which is both more stably transported and more easily measured than the RF phase of the heterodyne gauge. 4. It is able to operate in a resonant cavity, which improves precision and suppresses error due to misalignment. 5. It measures absolute distance with a minimum of added cost and complexity.

The TFG benefits from the convenience and effectiveness of Pound-Drever-Hall locking of a laser wavelength to a null of an interferometer. The Variable Frequency Source (VFS, Fig. 3) generates a laser signal of controllable optical frequency. The signal is phase modulated at a frequency $f_m$ and introduced into the interferometer. The signal emerging from the interferometer is also amplitude modulated at $f_m$, with a magnitude and sign that indicate the offset from null. Demodulation at $f_m$ yields an error signal that is fed back to the VFS, maintaining the lock.

In our current implementation, the Semiconductor Laser TFG (SL-TFG), the VFS comprises a tunable distributed feedback (DFB) semiconductor laser. Its optical frequency is locked to the cavity formed by the mirrors that define the measured length (see below). For all TFG's in one system, there is one reference laser, which is locked to a very stable cavity or to an atomic line. A portion of the light from each TFG's tunable laser is heterodyned against a portion of the reference laser light to provide the radio frequency output whose changes equal changes in optical frequency. The SR-POEM configuration has four SL-TFG's with closely-matched paths. The WEP observable is the difference of the averaged outputs from TMA-A and TMA-B. Because the measured paths are nearly equal, the reference laser frequency is quite uncritical.

The DFB laser tunes rapidly over a wide range, improving the ability of the TFG to follow vibration and to measure absolute distance. The SL-TFG uses rugged, fiber-coupled components, which reduces error due to thermal motion and reduces the instrument mass and volume. The DFB and associated fiber-coupled components were developed for the telecommunications industry, so they already meet stringent environmental and reliability requirements. DFB lasers have been flown as the metrology source for the Fourier Transform Spectrometer (FTS) in the Canadian Space Agency's Atmospheric Chemistry Experiment (ACE) on SCISAT-1, which was launched by NASA on Aug. 12, 2003.

The SR-POEM measurement will employ four conventional Fabry-Perot resonators. Each will employ a curved mirror at the TFG plate and a plane mirror at one of the cubes of one TMA. The optical beam will be injected through the curved mirror, which will have a partially-reflecting coating on the curved side and an anti-reflection coating on the flat side. Beam steering will be provided using an automated cavity alignment technique that is similar to that of Sampas and Anderson [Sampas 1990, Sampas and Anderson 1990]. Feedback to actuators on the beam-injection optics corrects the misalignments. Position and tilt alignment accuracy of 0.1 nrad Hz$^{-1/2}$ and 0.08 nm Hz$^{-1/2}$, respectively, were demonstrated by Sampas and Anderson, near the shot noise limit for the 160 μW power used. This is substantially better alignment than SR-POEM will require.

Must the plane of the mirrors on the TMA pass through the TMA CM? When the TMA temperature changes, thermal expansion causes an apparent motion proportional to the distance of the CM from the plane of the mirrors. However, the TMA temperature is expected to be constant to within $10^{-6}$ K over the ~1000 s mission. If the temperature variation with time were quadratic, $\delta T = (10^{-6} K)(t/t_m)^2$, where time runs from 0 to $t_m = 1000 s$ during the mission, and if the mirror were 3 cm above the CM and metered by aluminum, then the fictitious acceleration would be $1.5 \times 10^{-18}$ m s$^{-2}$ = $1.8 \times 10^{-19}$ g, which is completely negligible. This fictitious acceleration is constant in the payload frame and thus reversed in its effect on the successive measurements that contribute to the mission estimate of σ(η). Thus, the mirrors may be positioned on top of the cubes.

*2.4. Beam alignment and TMA rotation.*

Even if the TMA are not rotating with respect to the mean orientation of the payload, the ≈ $10^{-6}$ radian orientation variations of the payload will cause a misalignment of the optical cavity in the absence of an alignment servo. We address this at two levels. First, the alignment servo results quickly but results in "beam walk" on the curved mirror of ≈ 0.3 μm, which should be compared to the spot size, ω = 0.7 mm. Second, the required alignment correction is used to drive the hexapod motion system so as to bring the experiment back to the intended orientation. The alignment system is required for countering higher-frequency disturbances of low amplitude, while the hexapod motion system minimizes variations of the positions of the TMA with respect to their immediate surroundings.

Before each drop, the payload is oriented and the TMA-SS used to set the initial conditions, which ideally would be: the CM of the two TMA collocated; TMA CM collocated with payload CM; TMA oriented toward the TFG and not rotating with respect to the rest of the experiment.

The alignment of the TMA mirrors is achieved within the context of the automated cavity alignment technique mentioned above. The servo acts to keep the beam perpendicular to the flat mirror, and to keep the beam at the position on the curved mirror at which the mirror surface is parallel to the flat mirror. (The accuracy achieved by Sampas and Anderson was $10^{-10}$ radian Hz$^{-1/2}$, several orders of magnitude better than we require.) With the alignment system working, position changes on the curved mirror reflect changes in the angle of the flat mirror, with high accuracy, but with an unknown offset. The mirror's radius of curvature is ~0.3 m, so the sensitivity is ~$3\times10^{-6}$ radian/μm. We must measure this position to an accuracy of ~0.3 μm. To do this, we will place a beamsplitter in the beam just before it is injected into the cavity through the curved mirror. The diverted beam impinges on a quad cell (additional to the one required for the automated cavity alignment). The distances from beamsplitter to quad cell and to curved mirror will be the same, so that independent of beam angle, position changes on the quad cell and curved mirror will be equal.

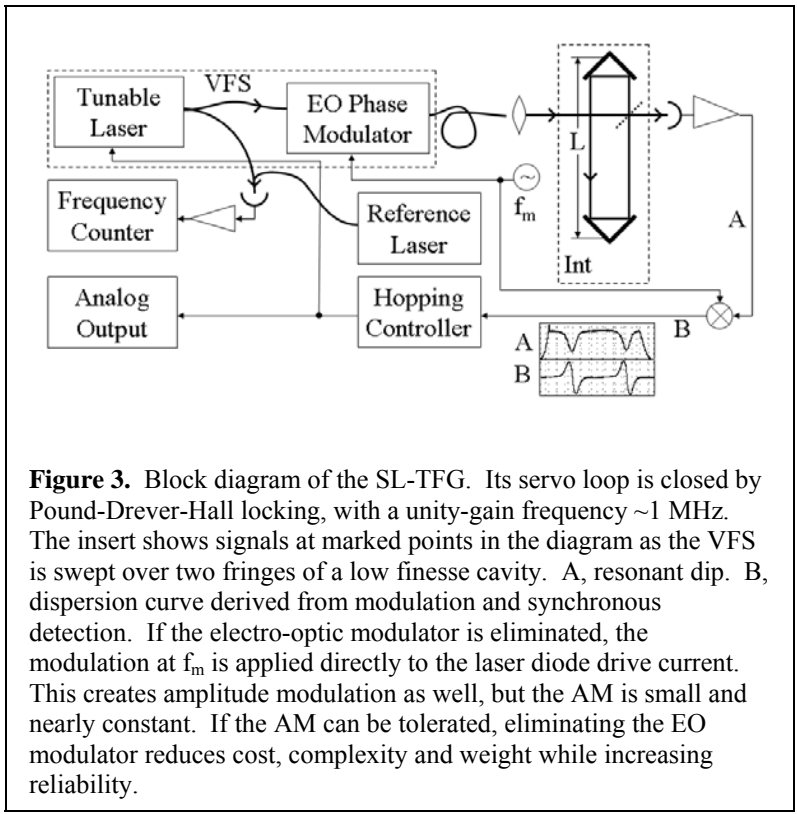

**Figure 3.** Block diagram of the SL-TFG. Its servo loop is closed by Pound-Drever-Hall locking, with a unity-gain frequency ~1 MHz. The insert shows signals at marked points in the diagram as the VFS is swept over two fringes of a low finesse cavity. A, resonant dip. B, dispersion curve derived from modulation and synchronous detection. If the electro-optic modulator is eliminated, the modulation at $f_m$ is applied directly to the laser diode drive current. This creates amplitude modulation as well, but the AM is small and nearly constant. If the AM can be tolerated, eliminating the EO modulator reduces cost, complexity and weight while increasing reliability.

The four quad cells must be calibrated so that when all four read zero (or near zero, with a stored offset for each), the four beams are parallel to the required ($2\times10^{-6}$ radian) accuracy. This calibration will be done on the ground by reflecting all four beams from a single flat placed in the plane that would contain the TMA mirrors, with the automated cavity alignment system in operation.

### 2.5. Capacitance gauge.

Each TMA will be surrounded by an electrode assembly (Fig. 4) of the capacitance gauge, which is part of the TMA-SS. The TMA-SS is capable of measuring and controlling all 6 degrees of freedom of each TMA. To measure, equal and opposite AC drive signals are applied to opposing pairs of electrodes. Different frequencies are used for each electrode pair. The potential on the TMA at the frequency for a particular electrode pair is zero when the TMA is centered, and linear in displacement for small displacements. The potential on the TMA is sensed by an amplifier connected to the sense electrodes. Control is by means of an additional signal applied to the appropriate parts of the electrode assembly at a frequency well separated from the frequencies used for measuring, and commensurate with the (1 ms) unit measuring interval. AC is used to keep from adding charge to the TMA or interacting with an existing net charge. During a WEP measurement, the capacitance gauge is disabled. During spacecraft inversion (end-for-end swaps), the TMA-SS controls all six degrees of freedom of the two TMA. This permits the TMA to move a long distance without touching the electrode assemblies (which could result in a change in the net TMA charge and possibly in sticking).

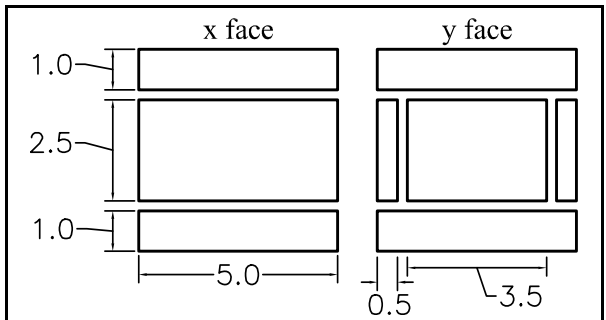

**Figure 4.** Plan of capacitance gauge electrodes. Dimensions are in cm. Interelectrode gaps are 0.25 cm. The long bars are drive electrodes and the large rectangle is the sense electrode. These electrodes are deposited on glass plates (ULE or Pyrex) that are later joined to form a box. Any significant open areas are filled with grounded shields, and the small remaining gaps are covered with material having a small conductivity which leaks off charge. A ground plane on the back of each plate covers the drive electrodes but not the sense electrode (because it would add capacitance and thus decrease sensitivity.) The design of the z face is not yet defined since it depends strongly on the design of the caging mechanism.

Based on preliminary results from the POEM capacitance gauge [Hickman 2007], we anticipate the SR-POEM capacitance gauge will have a sensitivity of $1\,\text{nm}/\sqrt{\text{Hz}}$ with a 30 mV drive signal.[3] The electric potential required to push the 1 kg TMA during payload inversions is < 1000 V.

Each TMA will acquire a charge at separation from its caging mechanism. It is convenient to refer to $\Delta V$, the TMA potential due to this charge when all surrounding surfaces are grounded, i.e., the potential picked up at liftoff. This charge could cause the TMA to accelerate, although perfect top-bottom symmetry of the TMA and housing would eliminate this effect. The TMA are nominally 4 mm from their top and bottom shields. For the top TMA surface (2 at 5.5 x 5.5 cm), a potential of 1 mV causes an acceleration of $1.8\ 10^{-16}$ g. Including the bottom surfaces, and taking the TMA to be 0.1 mm off center, the net acceleration is $0.17\ 10^{-16}$ g. This acceleration cancels in inversions. The capacitance gauge is used for initial centering. It measures the spacing difference directly. The TFG is used to precisely reproduce the TMA position after inversions. Thus, the change of TMA offset will be insignificant. We do not anticipate a source of additional charge after separation sufficient to change $\Delta V$ significantly during the short experiment.

In addition, there will be a spatially varying potential due to contaminants on the surface (patch effect). From studies of such variation over surfaces [Camp et al. 1991, Robertson et al. 2006], we expect $\Delta V$ <1 mV rms with respect to the mean. The attraction due to the mean surface potential of the top and bottom surfaces of a TMA and the corresponding capacitance plates, after the TMA is neutralized, can be removed by applying a separate potential to each plate to make the mean field zero. Of central importance is the stability of the surface potential. The best information we have about stability is from Fig. 5 of Robertson et al. [2006] and the corresponding text. From this, we conclude that change of force due to surface potential will not be a major contributor to the error budget since a fixed (small) force is cancelled well by the inversions.

To guard against unforeseen charging effects, we will test $\Delta V$ during flight, before and after the WEP measurements, and provide a capability for neutralizing the TMA before the WEP measurements are started. These tests can be done with higher atmospheric drag than is acceptable during WEP tests, so they can be performed at lower altitude. To sense $\Delta V$, we can apply equal and opposite potentials of 50 V to the top and bottom shields. Using the TFG, the acceleration due to $\Delta V=0.4$ mV can be measured ($\sigma = 0.03$ mV) in less than 1 s, half of that time with the external potential reversed. To neutralize this charge, we might illuminate the TMA and nearby shield with UV, and bias the electron transfer by applying a potetial to the shield. We estimate that measuring, neutralizing and remeasuring requires ~10 s per TMA. To prevent electric field changes near the TMA during the experiment, all surfaces capable of creating a field at the TMA will be at a well-defined potential, either a conductor connected to a well-defined potential or a resistive material bridging a gap between such conductors.

---

[3] The capacitance gauges would evolve from the ones being completed at the Rowland Institute at Harvard by our collaborator Winfield Hill.

The other key function of the capacitance gauges is to measure TMA velocity to allow correction for the Coriolis acceleration. If the payload is rotating at $2\pi$ per day, then to remove (differential) Coriolis error to an accuracy of $10^{-17}$ g in a run of 8 tosses requires knowing the difference of E-W velocity of the two TMA to an accuracy of 0.85 pm/s average for a run. This implies an accuracy of 2.4 pm/s in one 40 s toss. Since the Coriolis correction is linear in the E-W velocity, we only need to know the average velocity during a toss. Therefore, we will measure the velocity by applying a higher voltage for short periods both before and after the toss and require both positions to be measured with uncertainty under 70 pm. For example, by applying 1 V for 1 s we measure the velocity to the required accuracy. This measuring voltage on a single drive plate exerts a force that would move the TMA 20 pm. Again, there are several plates, but we can count on the applied force to be balanced to 1% for each opposing pair, yielding several displacements each of order 0.2 pm, a negligible perturbation to the average velocity during 40 s.

Above 800 km, the lateral component of drag acceleration will be under $10^{-12}$ g. This surprisingly low value comes from the low density of the remnant gas and from having a small horizontal velocity, $\approx 0.8$ km/s, far lower than orbital velocity, $\approx 8$ km/s (LEO). The spacecraft motion during 40 s due to this acceleration is only 8 nm, which is unimportant. The acceleration poses no additional problem.

### 2.6. Thermal design and analysis

In the short ($\approx 10^3$ s) period of free fall, it will not be possible to reach thermal equilibrium. Yet, thermal stability is an essential aspect of the WEP experiment. Our approach is to ensure long thermal time constants and, where necessary, to use low CTE materials. By flying around midnight, we avoid direct solar heating,[4] which would exceed other thermal perturbations and be highly directional in spacecraft coordinates. We note that the SR-POEM thermal problem is substantially easier than that for LISA in that the LISA payload assemblies must have an optical port to allow the entry of metrology laser beams. Below, we show that our quick experiment in a sounding rocket payload is in a thermally benign environment.

We seek to minimize two adverse thermal effects. First, temperature change within the precision measurement system can result in a direct measurement error. For example, it could cause a distortion of the TFG plate. Second, temperature change in any part of the payload can result in a change of the mass distribution and therefore a change in the accelerations of the TMA. Since a local mass generally couples differently to each TMA, a shift in the location of local mass will mimic a WEP signature. However, a slow position drift of local mass will tend to produce an effect that will cancel when the instrument orientation is reversed. At this stage in the concept development, we show more thermal isolation than is likely to be needed.

The four TFG are mounted on a highly stable plate, selected ULE glass with a coefficient of thermal expansion of $1\times10^{-8}$ / K. We require that the change of shape of the plate not introduce more than 0.01 pm into the WEP measurement. However, simple (cylindrical) bending or (spherical) doming will not matter to the extent that the TFG are symmetrically mounted, because the WEP signal is found by subtracting the average measurements of TMA-A from those of TMA-B. Because of the instrument inversions for successive WEP measurements, the offending shape change would need to have a component synchronous with the measurement cycles.

Figure 2 shows the instrument package. An aluminum vacuum chamber with ¼ inch wall is mounted to the payload tube through a thermally isolating active hexapod structure. The inside and outside of the chamber are gold plated giving those surfaces a low emissivity, $\varepsilon = 0.02$. Radiative exchange between the payload tube ($\varepsilon = 0.1$) and the chamber wall causes the chamber to approach the tube temperature with a time constant of $1.5 \times 10^5$ s. Inside the chamber, the most critical component is the TFG plate, which is a ULE disc of 20 cm diameter and 4 cm thickness with $\varepsilon \approx 1$. It is suspended from the metering tube by three titanium flexures. Radiative exchange between the chamber wall and the TFG plate causes the plate to approach the chamber wall temperature with a time constant of $5.5 \times 10^5$ s. The corresponding time constant for the much thinner (1 cm) metering tube is $1.4 \times 10^5$ s. These time constants were calculated without consideration of the beneficial effect of the mu-metal shield between the chamber and the ULE structure.

---

[4] Even at midnight, the Sun can be seen at the summer solstice (a few degrees above the horizon to the north) from 1200 km above Wallops Flight Facility.

What other paths could transport heat to the components inside the chamber? We have considered residual gas, which makes a negligible contribution at $10^{-8}$ Torr, and heat conduction through the support structure at the flange end of the chamber. We find that the total conductivity from flange to TMA plate is 34 mW/K, which gives a time constant of $10^5$ s. Not included in the heat leak inventory is the effect of the numerous wires that must penetrate the vacuum flange and terminate either at the capacitance gauge housings or on the TFG plate. None of these need carry much current, but many will be shielded. Heat transported through the TMA plate will contribute to the temperature difference along the z-axis across the TMA and the set of capacitance gauge plates. This temperature difference, in turn, contributes to forces on the TMA along the z axis from thermal radiation pressure and differential gas pressure. Both of these effects are small and are removed by the instrument inversion.

The connection from the payload tube to the outside of the vacuum flange can be made with thin-walled stainless steel tubing. As shown in Fig. 2, insulating pads (e.g., Conolite or Gorolite) can be used at both ends of the mounting structure. We estimate that this insulated structure will have a thermal conductivity similar to that calculated above.

What are the sources of varying temperature that could influence the experiment? The outer skin of the launch vehicle is discarded along with the nose cone early in the flight. The skin gets hot from the passage through the atmosphere and therefore will be insulated with 2 cm of urethane foam. The heat pulse lasts 100 s (before the hot skin is shed), which causes the payload tube to increase in temperature by 0.9 K.

Temperature differences in regions separated by 90 deg. on the payload tube are the most problematic since they could (depending on azimuthal phase) affect measurements of TMA-A and TMA-B differently. For this case, thermal diffusion provides no significant reduction, but payload inversion removes the effect from the final WEP determination. We do not know of any mechanism by which a small, slowly increasing temperature pattern could produce a false acceleration at the $10^{-17}$ g level.

We have investigated the thermal effect of the payload electronics on the experiment. We made a worst case assumption: The end of the payload tube would have the temperature rise, 0.04 K/s, seen typically in the hottest component, the communications transmitter. Figure 5 shows the temperature distribution along a 2 m tube. The payload is assumed to attach at 1 m from the heat source. After 1000 s, the instrument interface reaches 30 mK above ambient. This is small compared to the 900 mK from the hot skin and is not an issue.

Radiative cooling of the payload tube ($\varepsilon = 0.1$) will be at 4 mK / s. The payload inversions would expose one side of the payload tube to a warm Earth, asymmetrically reducing the radiative cooling during the 20 s slew. On the portion of the tube that most directly faces Earth, this could reduce the 2 K temperature change (over 500 s) by 10%. The TMA will be oriented such that two adjacent cubes, one from TMA-A and one from TMA-B, will be facing this "warming side" so that the effect will be equal on them, (and thus equal on the other pair). In this way, any possible systematic error due to this heating will be canceled. Again, the same payload inversion that causes this heating also removes its effect (say due to imperfect cancellation) on the WEP result.

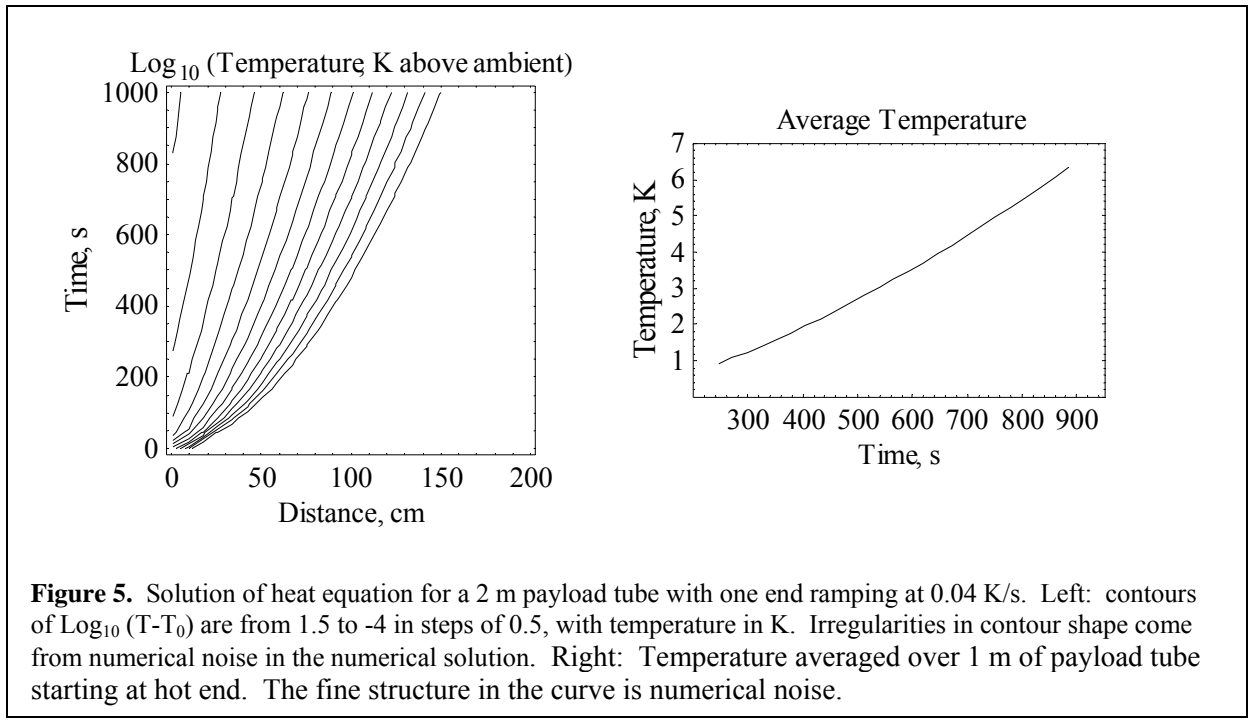

**Figure 5.** Solution of heat equation for a 2 m payload tube with one end ramping at 0.04 K/s. Left: contours of $Log_{10}$ (T-$T_0$) are from 1.5 to -4 in steps of 0.5, with temperature in K. Irregularities in contour shape come from numerical noise in the numerical solution. Right: Temperature averaged over 1 m of payload tube starting at hot end. The fine structure in the curve is numerical noise.

### 2.7. Local gravity.

During a series of measurements, the instrument z axis alternately points toward the nadir and the zenith. This payload inversion has the effect of reversing the WEP violating signal in the payload coordinate system. The gravity of the payload is unaffected by the inversion; it produces the same acceleration of the TMA in the instrument coordinate system independent of payload orientation.

However, the local mass distribution will not be stable through the experiment because the ACS gas will be consumed and because there will be temperature changes (and therefore size changes) in the payload components. The elastic response of the payload to a changing terrestrial gravity is negligible. The ACS gas will be stored in two connected tanks, each on the payload center line, on opposite sides of the experiment and approximately equidistant from the payload center of mass. Thus, the use of several kg of ACS gas will not significantly move the center of mass nor change the local gravity gradient by much.[5] However, as long as the tanks are on the center line, the change in local gravity will affect both TMA by the same amount and the payload inversion will mitigate any effect from off axis placement of a tank.

Figure 6 shows the effect on the WEP signal of a 10 micron displacement of 10 kg as a function of the radial and axial location of the mass. The nearby structure is ULE and has been shown to be at a stable temperature. It will not change the local gravity. Further away, the electronics packages are supported on aluminum shelves connected to the payload tube. During the measurement phase ($\approx$ 245 to 885 s) the last 100 cm of the tube will warm by $\approx$6 K, nearly linearly, resulting in an expansion of about 0.14 mm. The transmitter section, which is the major source of heat, has a mass of 36 kg, approximately symmetrically mounted around the payload axis to ensure dynamic balance. From Fig. 6, Right, a 5000 micron-kg displacement,[6] 1.5 m from the TMA and 5 cm off axis produces a $1.4 \times 10^{-18}$ g acceleration mimicking the WEP signal. The actual bias of the WEP estimate will be smaller because the far end of the payload tube will not be at the temperature of the hottest electronic component, the mass will likely be better centered, and the payload inversion will significantly cancel any effect.

---

[5] One kg at 1 m produces an acceleration of $6.8 \times 10^{-12}$ g. If a tank is 1 m from the TMA, 3 mm off axis and loses 1 kg of gas, it will cause a WEP mimicking acceleration of $10^{-18}$ g cos(2 β) (prior to cancellation by the inversions), where β is the azimuthal angle of the tank's lateral displacement from the axis of TMA-A.

[6] Almost all of the displacement mass is due to the displacement of the transmitter section. The contribution from the ACS section, which is located between the transmitter section and the instrument section, is neglected.

## 2.8. Gravity gradient

Near the Earth's surface, $dg/dh \approx 3 \times 10^{-7} g_0/m$ and $d^2g/dh^2 \approx 1.5 \times 10^{-13} g_0/m^2$, where $h$ is altitude and $g_0 = g(h)$ evaluated at $h = 0$. During a WEP measurement free-fall time Q, the TMA will move along the instrument z axis as a result of payload gravity, Earth's gravity gradient, and spacecraft acceleration due to low-thrust forces (drag, gas leaks, radiation pressure from a warm Earth, etc). They will therefore explore an inhomogeneous gravity field. In 40 s, the Earth's gravity gradient acting on TMA that are 1 cm from the horizontal plane through the payload CM will displace those TMA by 24 micron or, with proper initial velocity, 6 micron. We are considering adding a trim mass, so that we can reduce the CM offset to < 1mm.

Gravity changes due to portions of the payload that experience significant temperature change, so that their gravity field at the TMA changes, are treated above. An important source of gravity gradient is the shelf on which the TMA are caged during the ascent. Movement of the TMA with respect to the shelf by 3 nm changes the acceleration by 3 $10^{-17}$ g; the payload servo mentioned above is expected to reduce the relative motion. TFG measurements of TMA position with subpicometer resolution will support *a posteriori* correction, based on a moderately accurate model of the local gravity.

Since the TMA are not located at (but are quite close to) the center of mass of the payload, they accelerate away from (or toward) that point in proportion to their displacement from the CM multiplied by the Earth's gravity gradient tensor, plus a contribution from the payload's gravity. The gravity gradient tensor changes smoothly and predictably as the payload follows its free-fall trajectory; Earth's gravity field is precisely known, as will be the payload trajectory (*a posteriori*). For the payload close to its nominal (vertical) orientation, the gravity gradient produces very nearly the same TMA acceleration (in payload coordinates) independent of whether the instrument is nadir or zenith pointing. The differences, which come from the next term in the expansion of the Earth's field, are negligible. For the second order term to contribute an error in acceleration of $10^{-17} g$ would require the TMA CM's to be displaced vertically by about 1 cm. However, the CM's will coincide at the start of a measurement to better than 10 micron. Similarly, tilting the instrument by enough to displace the cubes of one TMA from the other pair by $\pm 1$ cm yields an additional differential acceleration of about $10^{-17} g$. We expect such tilts to be at least 1000 fold smaller.

## 3. Conclusion

We are developing a Galilean test of the WEP, to be conducted during the free-fall portion of a sounding rocket flight. The test of a single pair of substances is aimed at a measurement uncertainty of $\sigma(\eta) < 10^{-16}$ after averaging the results of eight separate drops. We have investigated sources of systematic error and find that all can be held to below the intended experiment accuracy. A detailed error budget is in preparation and will be published soon.

At the core of the mission design is the mitigation of systematic error, which is supported by the three stages of measurement differencing. First, the TMA positions are each measured by laser gauges with respect to the commoving instrument. These measurements contain the acceleration of the instrument and initial velocity errors, but the change over the 40 s drop period is at the micron level. Second, those measurements are differenced, removing the instrument acceleration. Third, the payload is inverted and the differential accelerations are differenced. This last step adds the WEP violating signals and subtracts the accelerations of the TMA due to fixed electrostatic, gas pressure and radiation pressure forces and to local gravity and the symmetric part of the Earth's gravity gradient. (The next, asymmetric term is both easily calculated and below the measurement threshold.) Ideally, at the start of a drop, the CM of each TMA should be at the CM of the payload. There will be an error in this setting, but the initial conditions for the drop will be highly reproducible because the key position (along the z axis) is sensed by the laser gauge.

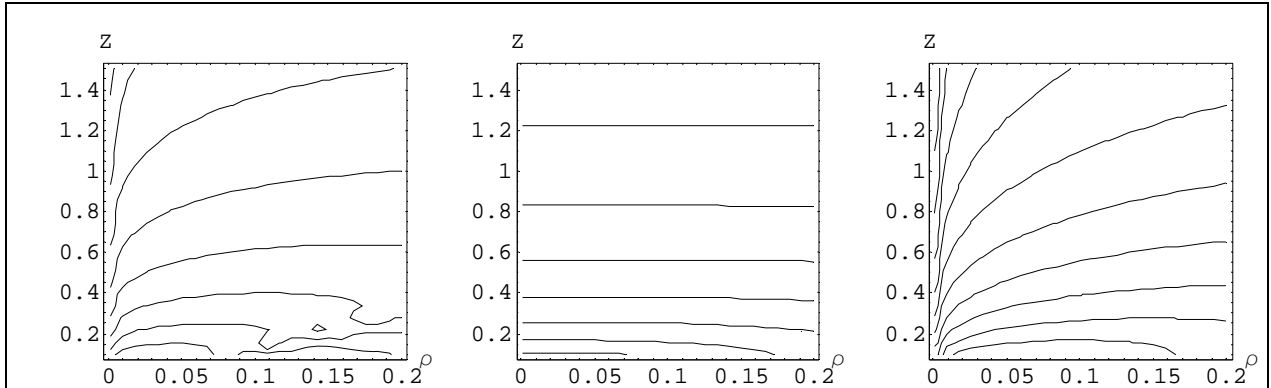

**Figure 6.** WEP mimicking acceleration in "g" from a displacement by 10 micron of a 10 kg mass as a function of the location of the mass with respect to the TMA CM. Position (in meters) shown is in the $\rho$ (radial) – z (axial) plane, with azimuthal orientation chosen to maximize the WEP mimicking acceleration (i.e., $\beta$ =0). Left: displacement in radial direction, contours from $10^{-13}$ to $10^{-20}$. Middle: displacement in azimuthal direction, contours from $10^{-16}$ to $10^{-22}$. Right: displacement in the z direction, contours from $10^{-13}$ to $10^{-22}$ .

Before we can perform a successful WEP experiment, we will need to develop an enhanced version of our TFG laser gauge, which will have a precision of 0.1 pm Hz$^{-1/2}$ and be self aligning. We will also need to demonstrate that we can release the TMA after launch and control them both during the inversion slews and in preparation for the drops. Methods for achieving these goals have been identified and work has been started on the demonstrations. The use of an active hexapod connection between the payload tube and the experiment chamber solves some problems related to the mitigation of systematic error. Although preliminary analysis showed no serious problems with the use of an active hexapod, this approach has not been subjected to detailed analysis and thus remains speculative .

## Acknowledgements

This work was supported in part by the NASA SMD through grants NNC04GB30G (past) and NNX08AO04G (present). Additional support came from the Smithsonian. We thank the staff at the Wallops Flight Facility for their generous contribution to our understanding of the capabilities and constraints of a sounding rocket flight.